\documentclass[pdflatex,sn-nature]{sn-jnl}

\usepackage{enumerate}
\usepackage{braket}
\usepackage{chemformula}
\usepackage{graphicx}%
\usepackage{multirow}%
\usepackage{amsmath,amssymb,amsfonts}%
\usepackage{amsthm}%
\usepackage{mathrsfs}%
\usepackage[title]{appendix}%
\usepackage{xcolor}%
\usepackage{textcomp}%
\usepackage{manyfoot}%
\usepackage{booktabs}%
\usepackage{algorithm}%
\usepackage{algorithmicx}%
\usepackage{algpseudocode}%
\usepackage{listings}%

\theoremstyle{thmstyleone}%

\theoremstyle{thmstyletwo}%

\theoremstyle{thmstylethree}%

\begin{document}

\title{
Electric field switching of chiral phonons
}

\author[1]{Michael Grimes}\email{michael.grimes@psi.ch}
\author[1]{Clifford J. Allington}
\author[1]{Hiroki Ueda}
\author[2]{Carl P. Romao}
\author[3]{Kurt Kummer}
\author[4]{Puneet Kaur}
\author[4]{Li-Shu Wang}
\author[4]{Yao-Wen Chang}
\author[4,5]{Jan-Chi Yang}
\author[1]{Shih-Wen Huang}
\author[1]{Urs Staub}\email{urs.staub@psi.ch}

\affil[1]{Center for Photon Science, Paul Scherrer Institut, 5232 Villigen, Switzerland}
\affil[2]{Czech Technical University in Prague, Prague 120 00, Czech Republic}
\affil[3]{ESRF - The European Synchrotron, 71 Avenue des Martyrs, F-38000 Grenoble, France}
\affil[4]{Department of Physics, National Cheng Kung University, Tainan 701, Taiwan}
\affil[5]{Center for Quantum Frontiers of Research \& Technology (QFort), National Cheng Kung University, Tainan 701401, Taiwan}


\abstract{
Lattice vibrations carrying angular momentum, known as chiral phonons, have emerged as a promising route to control and understand complex material properties, yet their deterministic manipulation remains largely unexplored. Here we demonstrate electric-field switching of phonon angular momentum in the technologically relevant ferroelectric \ch{BaTiO3}. Using circularly dichroic resonant inelastic X-ray scattering (CD-RIXS) at the oxygen K edge, we directly probe the phonon angular momentum and compare the measured dichroism with first-principles predictions of phonon-mode chirality. We find excellent agreement, revealing a momentum-dependent circular-dichroism contrast that exhibits a reversible gyroelectric effect, stable for at least 15 hours. Our results establish a robust mechanism for non-volatile control of chiral phonons and point towards new opportunities for phonon-based information and energy technologies.
}


\maketitle

More than a century after chirality was defined as the distinction between left- and right-handed structures, this concept continues to shape condensed matter physics in the spirit of Curie’s symmetry principle: it is asymmetry that creates the phenomenon \cite{Curie1894}.
Its influence spans from the enantioselective mechanisms that regulate molecular chemistry \cite{guijarro2009origin} to crystalline solids, where the absence of inversion symmetry gives rise to chiral order that couples spin, orbital, and lattice degrees of freedom. These couplings underpin a growing class of materials whose magnetic and electronic behaviours are governed by their inherent handedness.

Recently, attention has shifted from static structural chirality to lattice dynamics with the discovery of chiral phonons. These are defined as vibrational modes in which atoms break improper rotational symmetry \cite{Juraschek2025}. Such excitations carry quantised angular momentum and enable lattice-mediated angular-momentum transfer \cite{Park2020}. Arising in both chiral and non-centrosymmetric crystals \cite{Coh2023}, chiral phonons display handedness in their dynamic motion, imposing this handedness in their coupling to magnetic, electronic, and optical phenomena \cite{Bousquet_2025}.
This coupling has revealed a spectrum of emergent effects, including non-reciprocal phonon transport \cite{Aoki2019}, ultrafast magnetic switching \cite{Davies2024}, the phonon thermal Hall effect \cite{Uehara2022}, phonon magnetic moments \cite{Ren2021}, and spin–lattice angular-momentum exchange \cite{LevineChoi2024, Yao2025, Wu2025}. The angular momentum carried by phonons can be considered as an orbital magnetic moments through the chiral rotational motion of ionic charges, establishing one route for dynamical multi-ferroicity \cite{Juraschek2019} which has been proposed as a method for the ultrafast control of magnetisation and polarisation \cite{Basini2024,Davies2024}. The concept of phonon angular momentum (PAM) has thus emerged as a potential dynamical order parameter of chirality, analogous to magnetisation or ferroelectric polarisation \cite{Saparov2022,Juraschek2019}, offering a framework to quantify and, ultimately, control chiral states in solids \cite{Bousquet_2025}. Realising reversible control of PAM is therefore a critical step toward deterministic manipulation of lattice–driven phenomena governed by chiral symmetries.

Experimental confirmation of PAM at arbitrary points in reciprocal space achieved in the prototypical chiral crystal $\alpha$-quartz, using circularly polarised resonant inelastic x-ray scattering (RIXS) to reveal vibrational modes with well-defined angular momentum \cite{Ueda2023}. These results support long-standing theoretical predictions that chiral phonons originate from the non-degeneracy of left- and right-handed vibrational modes \cite{Vonsovskii1962, Levine1962}. Such non-degeneracy naturally arises in systems which break $\mathcal{PT}$ symmetry \cite{Coh2023}. This is realised either at general points in reciprocal space  for non-centrosymmetric crystals ($\mathcal{P}$) \cite{ueda2025_nc} or dynamic symmetry breaking in time-dependent responses \cite{Geilhufe2023,Chen2025a}.

\begin{figure*}[!ht]
    \centering
    \includegraphics[width=\linewidth]{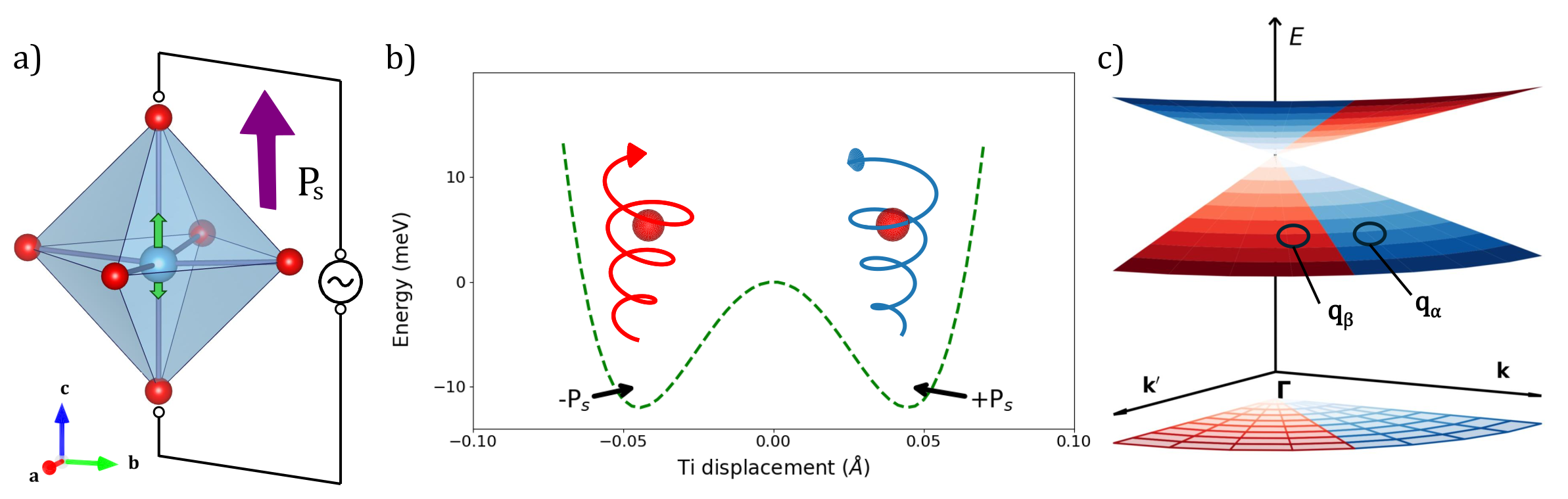}
    \caption{\textbf{Chiral phonons in a ferroelectric.} a) The bistable Ti displacements 
    relative to the crystal centre (oxygen octahedra) are the origin of displacive ferroelectricity, quantified by 
    the spontaneous electric polarisation, $\mathbf{P}_{s}$. By applying an electric field anti-parallel to the induced 
    polarisation, the vector direction of $\mathbf{P}_{s}$ is inverted. b) Depending on the direction of the Ti displacement (along the 
    c-axis of the crystal), the flipped polarity changes the handedness of the chiral phonons in the 
    material. Double-well taken from Cohen \cite{Cohen1992}. c) Chiral phonons are found in general 
    non-symmetric points ($\mathbf{q}_{\alpha}$) in the reciprocal space, where the handedness follows the 
    C\textsubscript{4V} symmetry of tetragonal BTO. Right and left handed phonons are non-degenerate in such 
    instances. Reflection about a high symmetry point (to $\mathbf{q}_{\beta}$) inverts the phonon angular 
    momentum.}
    \label{fig:intro}
\end{figure*}

While correlations between chiral lattice motion, phonon magnetism, and electronic topology are increasingly well established \cite{Hernandez2023,Dornes2019,Tauchert2022}, direct and reversible control of PAM remains largely unexplored. To address this challenge, we select a polar ferroelectric titanate that combines broken inversion symmetry with switchable polar order. A thin film of \ch{BaTiO3} (BTO) provides an ideal platform: its ferroelectric polarisation can be inverted \textit{in situ} by a small out-of-plane electric field, offering a controllable parameter of inversion-breaking symmetry \cite{Olaniyan2024, Cohen1992, Moseni2022}. Beyond its symmetry and ferroelectric properties, BTO is technologically relevant for low-power electronics, with applications including FeFETs, non-volatile memories, and neuromorphic architectures \cite{Yoo2025,Zhang2025_sci}. To maintain strong ferroelectric order while minimising strain \cite{Nordlander2020}, we employ free-standing epitaxial films with out-of-plane polarisation (Fig. \ref{fig:intro}). Using this system, we demonstrate non-volatile, reversible switching of phonon angular momentum through electric-field control of polarisation in ferroelectric BTO.

\subsection*{\ch{BaTiO3} Freestanding Films}

\begin{figure*}[!ht]
    \centering
    \includegraphics[width=0.95\linewidth]{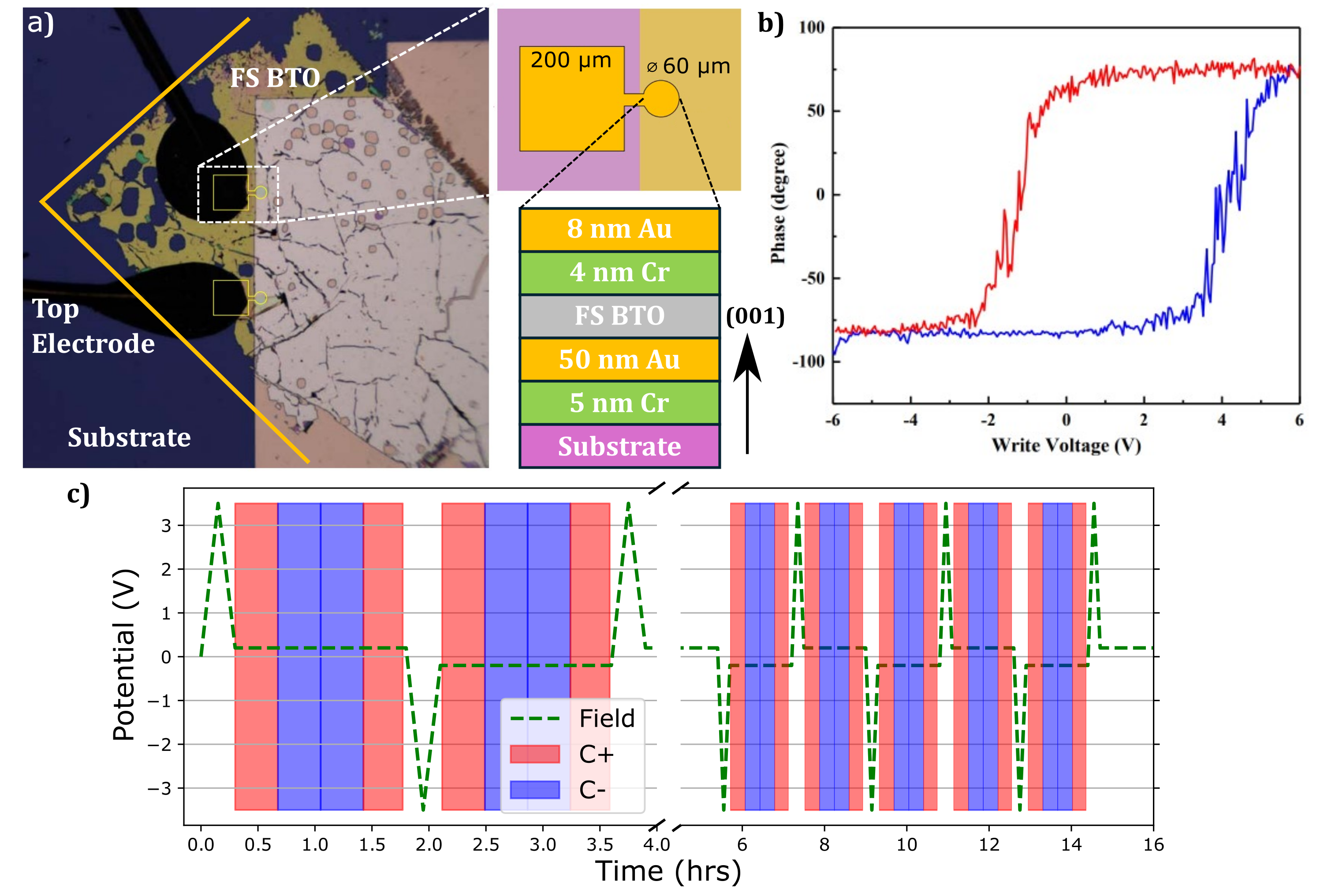}
    \caption{\textbf{Free standing films.} a) Freestanding films (FS) of ferroelectric BTO are 
    prepared and placed on Au$||$Cr bottom electrodes. Yellow lines indicate the a and b axes (equivalent) 
    where the c axis is out of plane. The top electrodes are deposited above the FS 
    film where the area corresponds to roughly to the x-ray spot size (50 microns long axis). b) The ferroelectric 
    nature and switching of the polarisation is confirmed using piezo force microscopy. The opening between increasing (blue) and decreasing (red) field yields a measure of coercivity of approx 3 V.
    c) Measurement scheme used during the experiment. The polarisation is switched by applying +(-) 
    3.5 V before a +(-) 0.2 V holding potential is maintained. In each ferroelectric state, RIXS is measured 
    with circularly right- and left-handed x-rays in a '+ - - +' pattern.}
    \label{fig:charac}
\end{figure*}

In order to study how direct control of ferroelectric polarisation can manifest as control of the handedness of chiral 
phonons, we address the ferro-electric \ch{BaTiO3} (BTO) crystal in its room temperature tetragonal phase using circularly polarised RIXS. Being polar, the crystal lacks inversion symmetry, resulting in phonons 
being non-degenerate and chiral at generic points in reciprocal space \cite{Coh2023,ueda2025_nc}. For structures described by the point group C\textsubscript{4v}, such as the ferroelectric phase of BTO, the projection of the phonon angular momenta in reciprocal space are reversed by inversion of the polarisation direction \cite{yang2025}. 
In this bulk phase, the \ch{Ti^4+} ion is not centred within the oxygen octahedron, producing a net electric polarization $P$ of the octahedra, \cite{Cohen1992}. By applying an electric field, the \ch{Ti^4+} 
can be displaced to the other stable site (double-well potential in Fig. \ref{fig:intro} b) thereby inducing a polarisation of opposite 
sign. Therefore, this system presents a perfect trial to test how the angular momentum (chirality) of certain phonon modes can be switched by an external electric field.To avoid epitaxially-induced strain and pinning of ferroelectric domains, samples of freestanding (FS) BTO are prepared \cite{Chiu2022}, see Methods and Figure \ref{fig:charac} a. This further allows the electric field to be applied across the plane of the sample. The properties of such perovskite layers are known to resemble bulk being free of inter-facial strain and substrate effects \cite{Leroy2025}. The intrinsic ferroelectricity of the FS-BTO was confirmed prior to the RIXS measurements, see Methods and Fig. \ref{fig:charac} b. The film orientation was confirmed by non-specular XRD prior to the synchrotron experiment (see Supplemental, Figure S.1 \& 2).

\subsection*{Evidence of chiral phonons in ferroelectric BTO.}

To demonstrate the reversal of angular momentum we use circularly polarised x-rays in RIXS, which intrinsically probes symmetry breaking via the transfer of angular momenta. This absorption-emission process has a cross-section 
defined by a second-rank tensor which is sensitive to electric monopole (charge), magnetic dipole (spin), and 
electric quadrupole (orbital asphericity) for the dominant X-ray scattering process, i.e., electric dipole-electric dipole 
(E1-E1) transitions \cite{Ament2011}. The X-ray Absorption Spectrum (XAS) around the O K-edge of the 
FS BTO film is presented in the Supplementary Materials (Figure S.3). For these x-ray energies, a $1s$ core electron is excited into a $2p$ valence shell.

The energy of the RIXS measurements was first refined to optimise the signal from the phonon excitation processes. The combined spectra can be found in Supplemental Figure S.5, where the strongest response was found for X-rays on the leading edge of the XAS spectra ($\approx$ 531.25 eV), roughly corresponding to the hybridised Ti-O orbitals \cite{Fan2019}. This energy was maintained for the 
remainder of the results presented in this paper. The phonon energies of BTO are limited to $\approx$ 100 
meV \cite{Ghosez1999}, where features at higher energies are presumed to be multi-phonon processes 
\cite{Bieniasz2022}.
One pressing issue that has been raised in recent RIXS studies is the influence 
of birefringence on the observed circular dichroism (CD) \cite{Nag2025} (see Supplemental Figure S.4). To mitigate this contribution, the field switching measurements were performed at normal incidence with beam parallel 
to the optical axis, c. 
Considering the above, the observed CD in RIXS originates from the excitation of O $2p$ electric quadrupoles, as CD is absent for 
isotropic charge scattering and BTO is non-magnetic. In fact, CD RIXS directly probes the rotation of the $2p$ orbitals with respect to the x-ray propagation via the dichroic contrast, $A_m $,  which is proportional to \cite{Ueda2023}
\begin{equation}
A_m \propto
\braket{ m \mid \mathbf{\epsilon}_c'^{*}
\begin{pmatrix}
0 & e^{-2i\phi}\\
e^{2i\phi} & 0
\end{pmatrix}
\mathbf{\epsilon}_c \mid 0 },
\label{eq:phon_over}
\end{equation}
where the eigenmodes, $\mathbf{\epsilon}_c$, are written in a circular basis (see Supplementary materials for derivation). As the transfer matrix which describes the polarisation dependence of RIXS only contains circular terms in the off-diagonal components (see Eq. \ref{eq:pol} in Methods), one can see how CD contrast arises from coupling to phonon modes with circular motion.
RIXS was performed at a general point in q-space so that we examine the CD contrast where the x-ray propagation 
is not aligned to high symmetry points of BTO (e.g. along the (110) glide plane). The outgoing angle is 
chosen to minimise the component along c, as the angular momentum is expected to be perpendicular to this axis 
(see phonon circular polarisation vectors in Supplementary Figure S.7). The respective RIXS spectra recorded for C+ and C- light are shown in Figure 
\ref{fig:chiral_phon} where $\mathbf{q}$ = (-0.12 , 0.20 , 0.42) r.l.u..
For a given polarisation direction, where one type of \ch{Ti^4+} displacement site is stabilised, the contrast between these two 
signals can be seen in Fig. \ref{fig:chiral_phon} b). Having discounted other sources of circular contrast in the RIXS signal, we therefore attribute this contrast to angular momentum transfer via the phonon system in ferroelectric BTO. 

The microscopic origin of this net angular momentum is due to the broken inversion symmetry being imposed in the 
force-constant applied to the ions in the crystal \cite{Coh2023}. For example, consider the interaction between the \ch{Ti^4+} 
and \ch{O^2-} ions in the structure shown in Fig. \ref{fig:intro} a. 
In general, the mechanical angular momentum of a phonon mode arises from the collective circular displacement $\mathbf{u}_{l \alpha}$ \cite{Juraschek2025};
\begin{equation}
    \mathbf{J} =   \sum_{l \alpha} m_{\alpha} \mathbf{u}_{l \alpha} \times \mathbf{u}_{l \alpha},
    \label{eq:am}
\end{equation}
where $m_{\alpha}$ is the atomic mass in the unit cell, $l$. Therefore, we expect non-degenerate chiral phonons with angular momentum projected in-plane (perpendicular to the polarisation axis).

We also note that the matrix elements which describe single-phonon coupling to the RIXS (in the absence of Coulomb 
interactions \cite{Johnston2010}) describe the projection of the oxygen contribution to the hybridised valence bands 
\cite{Devereaux2016}, as the x-ray energy corresponds to excitation of the Ti-O bonding environment \cite{Fan2019}. In short, the CD contrast measures the circular motion of 
the oxygen atoms as defined by Eq. \ref{eq:am}, where the reversed contrast indicates inversion of the circular rotation handedness (see Fig. \ref{fig:intro}).

Crucially, by symmetry arguments, it can be seen that switching between the opposite  
\ch{Ti^4+} displacements shown in Fig. \ref{fig:intro} a, the forces experienced by the oxygen ions 
would be inverted and angular momentum of the phonon modes should demonstrate reversal.

\subsection*{Reversible switching of chiral phonons} 

\begin{figure}[!ht]
    \centering
    \includegraphics[width=0.65\linewidth]{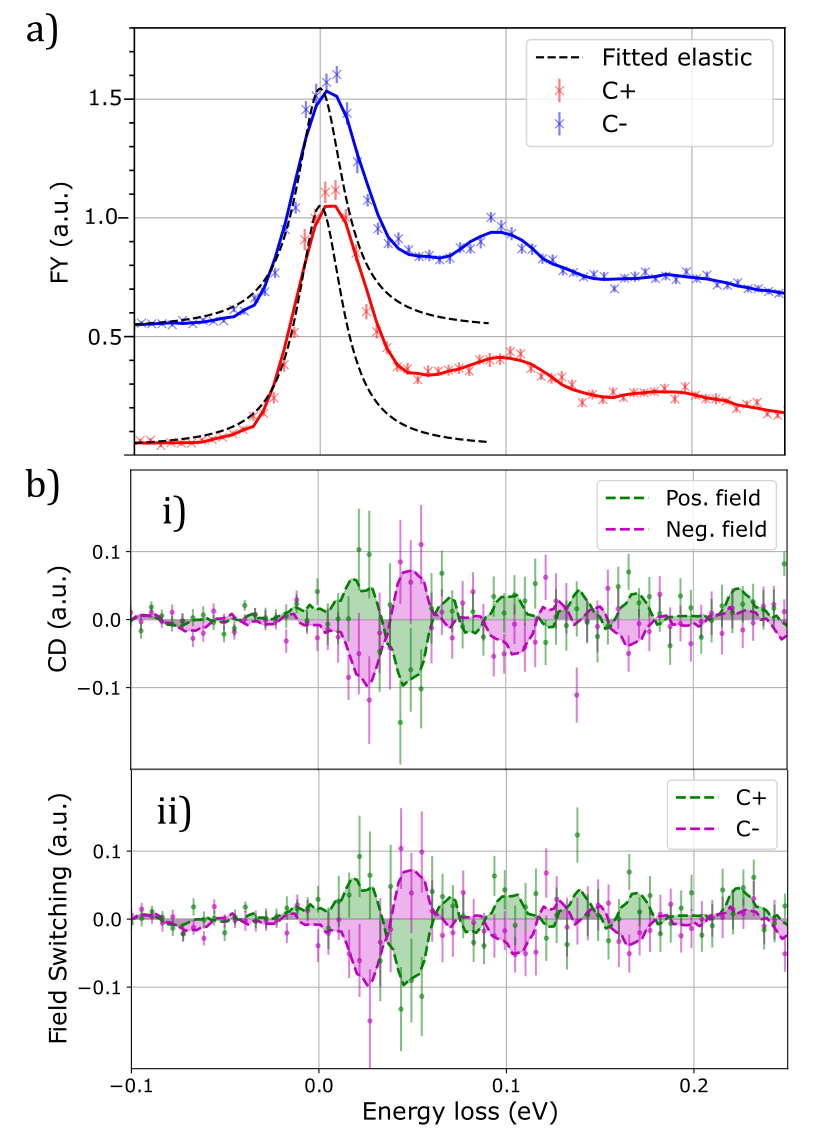}
    \caption{\textbf{Electric-field switching of chiral phonons measured with RIXS.} The RIXS spectra is recorded for two 
    X-ray polarisations under two different directions of applied electric field. a) Low energy 
    phonons are observed near the elastic line for x-rays energy corresponding to the near oxygen 
    K-edge (531.25 eV) when the sample is measured under positive electric field (0.2 V). The exact position of the elastic line is found by fitting the energy gain (negative energy loss) region with a Lorentzian curve, while Savitzky-Golay (SG) smoothing is applied to guide the eye. b) We confirm the switching of PAM by 
    measuring the CD in RIXS for a focused x-ray beam (50 microns). Upon reversing the applied 
    electric field, CD is inverted and remains stable for at least 15 hours. Likewise, the field-switching is inverted by 
    changing the left/right circular polarisation. Error bars are estimated from Poisson statistics 
    on re-binned data (by a factor 2), see Supplementary.}
    \label{fig:chiral_phon}
\end{figure}

By following the experimental procedure detailed in Methods and Fig. \ref{fig:charac} c we negate long-term signal drifts while switching the ferroelectric domain. To flip the domains, a potential of 
+(-) 3.5 V is temporally applied before measurements are performed with a holding potential of +(-) 0.2 V. This small voltage is applied to ensure the ferroelectric domains remain stable within the surface area of the contact pad during the duration of the measurements \cite{Yoo2025}. The resulting spectra under electric field switching are shown in Fig. \ref{fig:chiral_phon} for a given X-ray polarization. Having mitigated sources of birefringence, the circular dichroism 
shows switching under the application of external electric field.  As discussed above, CD contrast indicates reversal of the phonon handedness. The contrast 
is strongest for energy loss between 0 and 0.1 eV. The RIXS spectra in this region is a marker of the phonons 
where the signal arises due to electron-lattice coupling, where the phonon states are distinguished by their energies relative 
to the elastic line. The left-hand side (energy gain) of the elastic line is used to define the zero energy more precisely. Being sensitive only to thermally excited states (based on Bose-Einstein boson statistics) \cite{Ament2011}, a half Lorentzian peak (broadened delta function) describes the region reasonably well. An example of such fitting is seen as the dotted line in \ref{fig:chiral_phon}a.

An outstanding problem is the determination of individual phonon modes from the RIXS spectra. The features in the RIXS spectra are non-trivial to define, as reliable first principle calculations for electron-phonon coupling in RIXS cross section are not yet available. 
Further issues include the broadening of 
the phonon features as a function of the electron-phonon coupling strength \cite{Devereaux2016}. Under the Franck-Condon principle \cite{Ament2011}, the RIXS intensity can however be estimated as being proportional to the mode effective charge (see Supplementary section S4). With the knowledge of the phonon dispersion from DFT calculations (see Methods and Supplementary Figures S.6-8) and their calculated mode effective charges, we find perfect qualitative agreement with the measured phonon modes. For confirmation, by reducing the slit size of the exit beam (to reduce the angular spread of incident x-rays at the cost of total photons) we can perform a high-resolution RIXS scan to better determine the 
observed phonon modes. At the same values of $\mathbf{q}$ used in Fig. \ref{fig:chiral_phon}, RIXS at the oxygen K-edge was 
performed with a nominal energy resolution of 15 meV. From this analysis we can estimate the 
phonon contribution and highlight the phonon peaks observed at below 30 meV and between 30-60 meV.
The identification of the phonons modes is further refined by considering the intrinsic CD contrast, allowing us to isolate the contributions to phonons which impart a net angular momentum to the crystal \cite{Ueda2023}. This can be compared directly to the degree of chirality of the respective phonon modes as estimated from the DFT calculations on tetragonal BTO (see Methods). In Figure \ref{fig:chiral_phon}b, where we show the CD and field switching observed in the RIXS spectra, the strongest contrast is seen around 20, 40 and 100 meV which we attribute to angular momentum exchange with the predicted phonon modes at 11, 38, and 100 meV.

As shown in Fig. \ref{fig:chiral_phon} b, these regions exhibit switching of phonon chirality with respect to the direction of the electric poling field as well as to the inversion 
of the X-ray circular polarisation. These two observations provide direct evidence for the control of phonon angular momentum and chirality for modes within these regions.

\section*{Discussion \& Conclusions}

The origin of the phonon angular momentum, or the non degeneracy of chiral phonons, can be explained by considering the 
symmetry of the underlying crystal structure. For example, one can show that for a generic
non-symmetric point $\mathbf{q}$ in the Brillouin zone, a phonon will only possess net zero angular momentum 
when the material is invariant to both time-reversal and space inversion ($\mathcal{PT}$) symmetry \cite{Coh2023}. Furthermore, by 
considering an inversion operation, $\mathcal{P}$, the quantised angular momentum carried by the chiral phonon will 
be inverted when the anti-symmetric point, -$\mathbf{q}$ in the Brillouin zone is instead considered. 
Based on these considerations, the CD contrast ($A_m $ in Eq. \ref{eq:am}) will be limited to the non-symmetric 
points in reciprocal space. According to Curie's principle, chiral phonons in BTO must 
respect the symmetry of g-type C\textsubscript{4v} \cite{yang2025}. A sketch of the underlying symmetry can be 
found in Fig. \ref{fig:intro} c. As an aide, we note that the contrast observed in this work is inaccessible to optical measurements which are at a negligible $\vec{q}$ \cite{Coh2023}. 

This switching behaviour can be considered in terms of the well-established polarisation effects seen in chiral 
materials. In the initial studies of optical activity on chiral materials \cite{Bousquet_2025}, 
the Natural Optical Activity (NOA) arises from the spatial dispersion of the dielectric response. This is expressed in terms of a gyration tensor, $g_{ij}$,
\begin{equation}
    g_{ij} (\mathbf{q}, \omega) = \frac{1}{2} \epsilon_{kli} \lambda_{klj} (\mathbf{q}, \omega),
\end{equation} 
where $\epsilon_{kli}$ is the non-local dielectric response (which in our case depends also on $\mathbf{q}_k$),
and $\lambda$ is the momentum dependent NOA. When applying an electric field, the sign of optical 
rotation could be reversed in chiral ferroelectrics such as \ch{Pb5Ge3O11} in a phenomenon known as 
gyroelectricity \cite{Iwasaki1971}. This has further been observed to follow the dispersion laws of 
optical activity and refractive index \cite{Deliolanis2004}, having a linear relation which 
respects the crystal symmetry, $v$ \cite{Iwasaki1971}: 
\begin{equation}
    g_{ij} = \gamma_{ijv} P_s,
    \label{eq:gy_tens}
\end{equation}
where the gyration tensor, $\gamma_{ijv}$, is proportional to the spontaneous polarisation, $P_s$, of a ferroelectric. 
Here, we  observe a similar gyroelectric effect where the switching of the CD contrast seen in RIXS follows the same polarisation dependence. We also can attribute the microscopic origin of this contrast as the orbital moment of the chiral phonons modes. As the material is non-chiral, we expect the phenomena to be 
present only at non-symmetric points of reciprocal space. In agreement with the Curie principle, 
we can control the symmetry of the physical properties via direct control of the crystal symmetry; 
in this instance the electric field control of the inversion centre is manifested in the PAM of the chiral phonon.

\begin{figure*}[!ht]
    \centering
    \includegraphics[width=\linewidth]{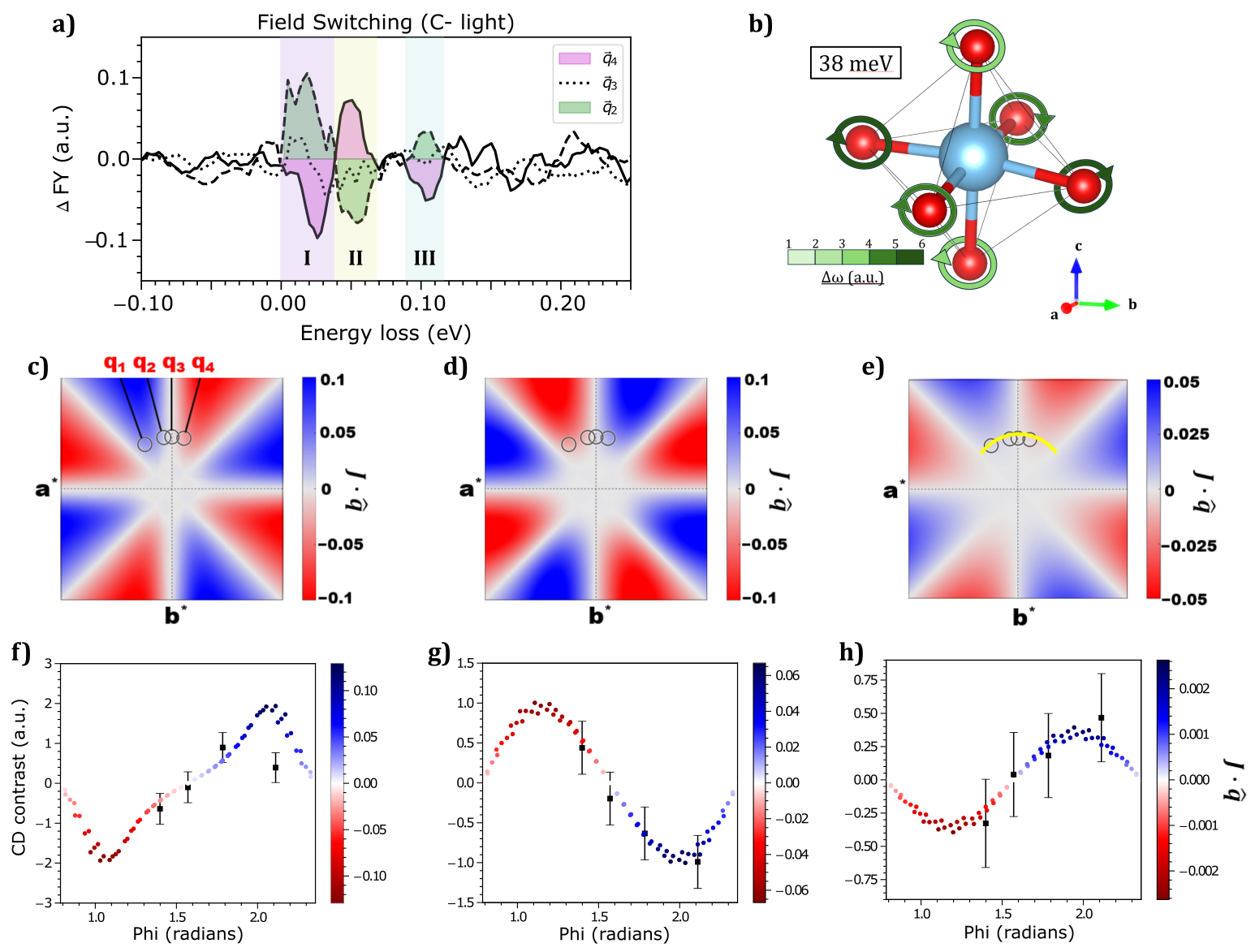}
    \caption{\textbf{Chiral phonons q-dependence.} a) Comparison of the field switching contrast (SG smoothed) seen in C- light for 
    different values of $\mathbf{q}$. The shaded regions correspond to the assumed chiral phonons. Reversal of contrast is observed for inversions of $\mathbf{q}$ ($\vec{q}_2$ vs $\vec{q}_4$). b) Example 
    of the circular motion (AM projected in-plane) of the oxygen atoms at $\vec{q}_2$ computed from the DFT simulations of the tetragonal BTO (videos are available in the Supplementary materials). c-e) Computed values 
    of $\mathbf{J} \cdot \mathbf{q}$ where $q_z$ is fixed at 0.42 \AA \textsuperscript{-1} for 3 phonon modes which have significant PAM, where 
    the energies are 11, 38 and 100 meV (associated with regions I, II, and III above). The contrast follows the g-type C\textsubscript{4v} symmetry \cite{yang2025} of the BTO structure. 
    f-h) Integrated CD contrast for the shaded regions (I, II, and III) in a) as a function of $\phi$. These 
    correspond to points shown along the yellow line in e). The variation in the measured contrast matches the calculated values well for the 
    3 phonon modes across the directions measured.}
    \label{fig:q_dep}
\end{figure*}

A key component of such RIXS studies which examine the phonon spectra
is the previously discussed electron-phonon coupling, where the contrast emerges from the transfer of angular momentum to the phonon system. Due to the $C_{4v}$ symmetry 
of the unit cell, the signal of the phonon contributions should exhibit a similar four-fold symmetry with respect to the experimental geometry, effectively describing a gyration tensor as defined in Eq. \ref{eq:gy_tens}. For a given phonon angular momentum $\mathbf{J}$ (see Eq. \ref{eq:am}), we would expect 
the contrast to be captured in the chiral projection 
parameter $\mathbf{J} \cdot \mathbf{q}$, known as the phonon helicity only being non-zero for truly chiral phonons (those which break improper rotational symmetry) \cite{Juraschek2025}. Crucially, along certain crystallographic
directions, we would therefore expect the contrast to invert for small rotations 
of the sample. This is highlighted in Figure \ref{fig:q_dep}, where the chiral 
projection from the three lowest lying chiral phonons are computed from DFT calculations 
of the phonon frequencies based on the BTO tetragonal structure (see Methods).

\begin{table}[!t]
\caption{RIXS at the oxygen K-edge was performed at the following $\mathbf{q}$ values,
for a fixed x-ray energy of 531.25 eV.}
\label{tab:q_dep}

\begin{tabular}{||c | c c c||}
    \hline
    & q\textsubscript{x} (\AA $^ {-1}$) & 
        q\textsubscript{y} (\AA $^ {-1}$) & 
        q\textsubscript{z} (\AA $^ {-1}$) \\ 
    \hline\hline
    $\vec{q}_1$ & -0.12 & 0.20 & 0.42 \\ 
    \hline
    $\vec{q}_2$ & -0.05 & 0.23 & 0.42 \\
    \hline
    $\vec{q}_3$ & 0.00 & 0.24 & 0.42 \\
    \hline
    $\vec{q}_4$ & 0.04 & 0.23 & 0.42 \\
    \hline

\end{tabular}
\end{table}

In Tab. \ref{tab:q_dep} we list the four directions of $\mathbf{q}$ where RIXS was measured 
during this experiment. 
This was achieved by rotating the sample around the surface normal, $\phi$, while
other diffractometer and spectrometer angles are held constant to vary the in-plane component of $\mathbf{q}$ 
($q_x$ and $q_y$). This ensures the projection of the PAM exchange with the circularly
polarised x-rays will only follow this component, thereby confirming the origin of the 
CD contrast as being the predicted circular rotation of the oxygen atoms (being sensitive only to out-of plane ion 
displacements). This relation was observed experimentally for the two chiral phonons in the BTO 
system with the strongest CD contrast, presumed to be those at 11 meV, and 38 meV (see Supplementary 
for the BTO dispersion). In Figure \ref{fig:q_dep}, the inversion of the field switching is seen in 
the signals when comparing the RIXS spectra from $\vec{q_2}$ to that of $\vec{q_4}$, where the chiral 
projections have (almost) equal magnitude but opposite sign. An example of the improper rotational motion of the oxygen atoms predicted from the DFT calculations is shown in Fig. \ref{fig:q_dep}b) for the 38 meV phonon along the $\vec{q}_2$ direction. The angular momentum ($\omega$) is estimated as the maximal area of the ellipse the oxygen atom traverses such that the normal is projected in-plane. This is proportional to $\mathbf{J}$ as described by Eq. \ref{eq:am} since inversion symmetry is broken in the out-of-plane direction. In Fig. \ref{fig:q_dep}f-h, we compare the magnitude of the CD-RIXS contrast in regions I, II, and III of Fig. \ref{fig:q_dep}a to the computed $\mathbf{J} \cdot \mathbf{q}$ as a function of $\phi$. When scaled accordingly, the observed values follow those predicted within experimental error. Furthermore, we note in Fig. \ref{fig:q_dep}c 
that the CD contrast is largest near $\vec{q_1}$ where the momentum transfer is most parallel to $\mathbf{J}$. The outlier at this point in Fig \ref{fig:q_dep}f arises due to the smearing of the signal across several low-lying phonon modes (energy resolution $\approx$ 30 meV). The weighted vector projection of these convolved modes is available in Supplementary Figure S.11.

This confirms that the CD-RIXS contrast can be inverted by measuring along an inverted $\mathbf{q}$, 
verifying that the relationship follows the crystallographic symmetry and the PAM
must arise from the broken inversion symmetry of the displacive ferroelectric phase.
The present work demonstrates deterministic inversion of phonon angular momentum in a technologically relevant perovskite over at least 15 hours period. Previous studies of chiral phonons has been dedicated to the examination of their coupling to a range of secondary phenomena, including the phonon Hall effect \cite{Park2020}, lattice dynamics \cite{Saparov2022}, emergent magnetic fields \cite{Xiong2022}, and spin polarisation \cite{Fransson2023}. 
This form of gyro-electric control facilitates active manipulation of chiral phonon populations, thereby establishing an alternative route for phononics in which angular momentum, rather than frequency alone, becomes a tunable degree of freedom. 
As polarisation is an order parameter in general, this approach allows a much more complete manipulation of phonon angular momentum, offering unprecedented control in technologically mature ferroelectrics. This opens the prospect for integrating THz- or optically-driven, angular-momentum-carrying phonons (coupled to adjacent magnetic layers) to realise on-chip phonon–magnon reservoirs for neuromorphic computing \cite{Zhang2025_sci,Yaremkevich2023}, bridging ultrafast lattice dynamics with non-volatile synaptic states.

\section*{Methods} \label{sec:methods}

\subsection*{Growth of BTO}

BTO thin films were fabricated by pulsed laser deposition (PLD) using a KrF excimer laser with a wavelength of 248 nm. To enable the preparation of freestanding BTO, epitaxial BTO/LSMO heterostructures were first grown on (001)-oriented \ch{SrTiO_3} substrates. The LSMO sacrificial layers were deposited at 700 $^{\circ}$C under an oxygen partial pressure of 100 mTorr, followed by the growth of BTO layers with various thicknesses at 725 $^{\circ}$C and the same oxygen pressure. For both layers, a laser energy of 200 mJ and a repetition rate of 10 Hz were employed. After deposition, the samples were cooled to room temperature in an oxygen atmosphere of 760 Torr to suppress oxygen vacancy formation and preserve the crystalline quality of the heterostructures.

\subsection*{Fabrication of the FS-BTO polarization switching device architecture}

To obtain freestanding BTO, the as-grown BTO/LSMO heterostructures were subjected to a chemical etching process by immersing the samples in a diluted hydrochloric acid (HCl) solution, which selectively removed the LSMO sacrificial layer and separated the freestanding BTO membranes from the substrates. \\

The device structure for FS-BTO polarization switching was fabricated by combining standard photolithography, thermal evaporation, and the transfer of freestanding BTO membranes. First, S1813 photoresist was spin-coated onto Si substrates coated with a 300 nm \ch{SiO_2} layer and baked at 135 $^{\circ}$C for 60 s. LED lithography was then employed to define the bottom-electrode geometry. Subsequently, a 2 × 2 mm\textsuperscript{2} Au/Cr bottom electrode (50/5 nm in thickness) was deposited by thermal evaporation. The freestanding BTO membranes were then transferred onto the bottom electrodes.\\

To enable polarization switching of the freestanding BTO membranes, a specially designed jigsaw-puzzle-like top electrode was fabricated using the same LED lithography and thermal evaporation procedures. The top electrode consisted of 4 nm Cr and 8 nm Au, chosen to ensure sufficient X-ray transmission for CD-RIXS measurements.

\subsection*{Electrical tests}

The sample is confirmed to be ferroelectric in nature by conducting IV scans with an oscilloscope prior to the experiment. 
Using the protocol described by Fina \textit{et al.} \cite{Fina2011}, we confirm that the ferroelectric domains begin to 
switch around applied fields of 3 V. This was further confirmed using piezo-force microscopy (PFM) performed on the FS 
films Fig. \ref{fig:charac} b. 

\subsection*{RIXS}

RIXS measurements were performed at the RIXS end-station of Beamline ID32 at the ESRF, Grenoble, France \cite{BROOKES2018175}. We tuned the photon energy to around the O K edge. The energy resolution was
estimated as $\approx$ 30  (15) meV full width at half-maximum from the elastic peak of a carbon tape for field switching (high-resolution) scans. A
BTO freestanding film with a single ferroelectric domain state and the largest face perpendicular to c
in the tetragonal setting was prepared. The goniometer of the RIXS end-station
allows us to access different momentum points during the experiment. The error bars in an
RIXS spectrum are the standard deviation of individual scans. The X-ray absorption spectrum
was obtained by the total fluorescence yield before/during the RIXS measurements.

\subsubsection*{Birefringence in RIXS}

Based on the transfer matrix formalism 
based on Jones calculus, the polarisation matrix may be written as \cite{Jones_48};
\begin{equation}
\Pi =
 \frac{1}{2}
 \begin{pmatrix} 
    1 + S_1 & S_2 -i S_3 \\
    S_2 + iS_3 & 1 - S_1
 \end{pmatrix},
 \label{eq:pol}
\end{equation}
where $S_3 = 1 \Rightarrow$ C+ and $S_3 = -1 \Rightarrow$ C- x-rays. Therefore, depending on the incidence 
angle and the relative cross section with the in-plane and out-of-plane refractive indices, the depth-dependent 
polarisation will change \cite{Nag2025}. This effect is shown in the Supplementary (Figure S.4). Strong circular dichroism (CD) is observed 
which masks the CD contrast due to the chiral phonon interaction. 

\subsubsection*{Ferroelectric switching in RIXS}

By applying write (3.5 V) and hold (0.2 V) voltages, we can switch and maintain the 
ferroelectric domain within the region of the top electrode. The measurement procedure is illustrated in 
Fig. \ref{fig:charac} c. CD-RIXS is measured in each ferroelectric state under the holding
potential by switching between C+ and C- light in a '+ - - +' fashion. The out-of-plane potential is
reversed and the procedure is performed as before.

\subsection*{DFT}

Density functional perturbation theory calculations of the phonon frequencies
and eigenvectors of BTO were performed using the Abinit software package (v. 10) \cite{Verstraete2025}. The calculations
used the PBE GGA exchange–correlation functional with the vdw-DFT- D3(BJ) dispersion correction. The PAW method was used with a plane-wave basis set cutoff energy of 140 Ha within the PAW
spheres and 28 Ha without. PAW basis sets were used as received from the Abinit library. A 6 x 6 x 6
Monkhorst–Pack grid was used to sample both k-points and q-points. The k-point grid spacing and
plane-wave basis set cut-off energy were chosen following convergence studies, with the convergence
criterion being 1\% in pressure. Prior to the phonon calculation, the structure was relaxed to an internal
pressure of 70 kPa, resulting in hexagonal lattice constants of a = 3.99 \AA \ and c = 4.11 \AA, in reasonable agreement  with experimental values (a = 4.00 \AA \ and c = 4.03 \AA) \cite{Culbertson2020}. \\

\backmatter
\bmhead{Data availability}

Experimental data acquired at the ESRF can be found in the ESRF data portal at DOI:\href{https://doi.esrf.fr/10.15151/ESRF-ES-2166894199}{10.15151/ESRF-ES-2166894199} and will be available upon publication. Computational data are available from DOI:\href{https://zenodo.org/records/17899320}{10.5281/zenodo.17899320}\\

\bmhead{Supplementary Information}

Supplementary information is available for this paper.\\

\bmhead{Correspondence and requests for materials}

Correspondence should be addressed to M. G. or U.S. (michael.grimes@psi.ch, urs.staub@psi.ch).\\

\bmhead{Acknowledgements}

The authors are grateful to ESRF synchrotron for allocating beamtime \cite{bto_grimes} under proposal number HC-6116, 
and to the ID32 beamline staff for the excellent support during the measurements. J.-C.Y. 
acknowledges the financial support from the National Science and Technology Council 
(NSTC), Taiwan, under grant nos.
112-2112-M-006-020-MY3, 114-2124-M-006-003, and 115-2923-M-006-001-MY4. M. G. and C. J. A.  acknowledge
funding from the Swiss National Science Foundation (SNSF) under grant number 10.000.807. 
C.P.R. acknowledges support from the project FerrMion of the Ministry of Education, Youth and Sports, Czech Republic, co-funded by the European Union (CZ.02.01.01/00/22\_008/0004591). 
Computational resources were provided by the state of Baden-Württemberg through bwHPC and the Deutsche Forschungsgemeinschaft through grant no INST 40/467-1 FUGG (JUSTUS cluster).\\

\bmhead{Author Contributions}
U.S., and H.U. conceived the idea and designed the experiments. 
J-C. Y was responsible for the sample fabrication. 
P.K. and L-S.W. performed the sample growth. Y-W. C. transferred the sample and fabricated the electrical contacts.
S-H. W. pre-characterised the sample at the ADDAMS (X04SA) of the SLS, PSI Villigen.
C. A., H. U., U.S., and M. G. performed the RIXS beamtime at ID32, with assistance from K.K.
C. R. performed and analysed the DFT calculations.
M.G. analysed the results and composed the paper.


\begin{thebibliography}{10}
\expandafter\ifx\csname url\endcsname\relax
  \def\url#1{\burl{#1}}\fi
\expandafter\ifx\csname urlprefix\endcsname\relax\def\urlprefix{URL }\fi
\providecommand{\bibinfo}[2]{#2}
\providecommand{\eprint}[2][]{\url{#2}}
\providecommand{\doi}[1]{\url{https://doi.org/#1}}
\bibcommenthead

\bibitem{Curie1894}
\bibinfo{author}{{Curie, P.}}
\newblock \bibinfo{title}{Sur la symétrie dans les phénomènes physiques, symétrie d'un champ électrique et d'un champ magnétique}.
\newblock \emph{\bibinfo{journal}{J. Phys. Theor. Appl.}} \textbf{\bibinfo{volume}{3}}, \bibinfo{pages}{393--415} (\bibinfo{year}{1894}).
\newblock \urlprefix\url{https://doi.org/10.1051/jphystap:018940030039300}.

\bibitem{guijarro2009origin}
\bibinfo{author}{Guijarro, A.} \& \bibinfo{author}{Yus, M.}
\newblock \emph{\bibinfo{title}{The Origin of Chirality in the Molecules of Life: A Revision from Awareness to the Current Theories and Perspectives of this Unsolved Problem}}  (\bibinfo{publisher}{The Royal Society of Chemistry}, \bibinfo{address}{London}, \bibinfo{year}{2008}).
\newblock \urlprefix\url{https://doi.org/10.1039/9781847558756}.

\bibitem{Juraschek2025}
\bibinfo{author}{Juraschek, D.~M.} \emph{et~al.}
\newblock \bibinfo{title}{Chiral phonons}.
\newblock \emph{\bibinfo{journal}{Nature Physics}} \textbf{\bibinfo{volume}{21}}, \bibinfo{pages}{1532--1540} (\bibinfo{year}{2025}).
\newblock \urlprefix\url{https://doi.org/10.1038/s41567-025-03001-9}.

\bibitem{Park2020}
\bibinfo{author}{Park, S.} \& \bibinfo{author}{Yang, B.~J.}
\newblock \bibinfo{title}{Phonon angular momentum hall effect}.
\newblock \emph{\bibinfo{journal}{Nano Lett.}} \textbf{\bibinfo{volume}{20}}, \bibinfo{pages}{7694--7699} (\bibinfo{year}{2020}).
\newblock \urlprefix\url{https://doi.org/10.1021/acs.nanolett.0c03220}.

\bibitem{Coh2023}
\bibinfo{author}{Coh, S.}
\newblock \bibinfo{title}{Classification of materials with phonon angular momentum and microscopic origin of angular momentum}.
\newblock \emph{\bibinfo{journal}{Phys. Rev. B}} \textbf{\bibinfo{volume}{108}}, \bibinfo{pages}{134307} (\bibinfo{year}{2023}).
\newblock \urlprefix\url{https://doi.org/10.1103/PhysRevB.108.134307}.

\bibitem{Bousquet_2025}
\bibinfo{author}{Bousquet, E.} \emph{et~al.}
\newblock \bibinfo{title}{Structural chirality and related properties in periodic inorganic solids: review and perspectives}.
\newblock \emph{\bibinfo{journal}{Journal of Physics: Condensed Matter}} \textbf{\bibinfo{volume}{37}}, \bibinfo{pages}{163004} (\bibinfo{year}{2025}).
\newblock \urlprefix\url{https://doi.org/10.1088/1361-648X/adb674}.

\bibitem{Aoki2019}
\bibinfo{author}{Aoki, R.}, \bibinfo{author}{Kousaka, Y.} \& \bibinfo{author}{Togawa, Y.}
\newblock \bibinfo{title}{Anomalous nonreciprocal electrical transport on chiral magnetic order}.
\newblock \emph{\bibinfo{journal}{Phys. Rev. Lett.}} \textbf{\bibinfo{volume}{122}}, \bibinfo{pages}{057206} (\bibinfo{year}{2019}).
\newblock \urlprefix\url{https://link.aps.org/doi/10.1103/PhysRevLett.122.057206}.

\bibitem{Davies2024}
\bibinfo{author}{Davies, C.~S.} \emph{et~al.}
\newblock \bibinfo{title}{Phononic switching of magnetization by the ultrafast barnett effect}.
\newblock \emph{\bibinfo{journal}{Nature}} \textbf{\bibinfo{volume}{628}}, \bibinfo{pages}{540--544} (\bibinfo{year}{2024}).
\newblock \urlprefix\url{https://doi.org/10.1038/s41586-024-07200-x}.

\bibitem{Uehara2022}
\bibinfo{author}{Uehara, T.}, \bibinfo{author}{Ohtsuki, T.}, \bibinfo{author}{Udagawa, M.}, \bibinfo{author}{Nakatsuji, S.} \& \bibinfo{author}{Machida, Y.}
\newblock \bibinfo{title}{Phonon thermal hall effect in a metallic spin ice}.
\newblock \emph{\bibinfo{journal}{Nature Communications}} \textbf{\bibinfo{volume}{13}}, \bibinfo{pages}{4604} (\bibinfo{year}{2022}).
\newblock \urlprefix\url{https://doi.org/10.1038/s41467-022-32375-0}.

\bibitem{Ren2021}
\bibinfo{author}{Ren, Y.}, \bibinfo{author}{Xiao, C.}, \bibinfo{author}{Saparov, D.} \& \bibinfo{author}{Niu, Q.}
\newblock \bibinfo{title}{Phonon magnetic moment from electronic topological magnetization}.
\newblock \emph{\bibinfo{journal}{Phys. Rev. Lett.}} \textbf{\bibinfo{volume}{127}}, \bibinfo{pages}{186403} (\bibinfo{year}{2021}).
\newblock \urlprefix\url{https://doi.org/10.1103/PhysRevLett.127.186403}.

\bibitem{LevineChoi2024}
\bibinfo{author}{Choi, I.~H.} \emph{et~al.}
\newblock \bibinfo{title}{Real-time dynamics of angular momentum transfer from spin to acoustic chiral phonon in oxide heterostructures}.
\newblock \emph{\bibinfo{journal}{Nature Nanotechnology}} \textbf{\bibinfo{volume}{19}}, \bibinfo{pages}{1277--1282} (\bibinfo{year}{2024}).
\newblock \urlprefix\url{https://doi.org/10.1038/s41565-024-01719-w}.

\bibitem{Yao2025}
\bibinfo{author}{Yao, D.} \& \bibinfo{author}{Murakami, S.}
\newblock \bibinfo{title}{Theory of spin magnetization driven by chiral phonons}.
\newblock \emph{\bibinfo{journal}{Phys. Rev. B}} \textbf{\bibinfo{volume}{111}}, \bibinfo{pages}{134414} (\bibinfo{year}{2025}).
\newblock \urlprefix\url{https://doi.org/10.1103/PhysRevB.111.134414}.

\bibitem{Wu2025}
\bibinfo{author}{Wu, F.} \emph{et~al.}
\newblock \bibinfo{title}{Magnetic switching of phonon angular momentum in a ferrimagnetic insulator}.
\newblock \emph{\bibinfo{journal}{Phys. Rev. Lett.}} \textbf{\bibinfo{volume}{134}}, \bibinfo{pages}{236701} (\bibinfo{year}{2025}).
\newblock \urlprefix\url{https://doi.org/10.1103/gcgl-9sbb}.

\bibitem{Juraschek2019}
\bibinfo{author}{Juraschek, D.~M.} \& \bibinfo{author}{Spaldin, N.~A.}
\newblock \bibinfo{title}{Orbital magnetic moments of phonons}.
\newblock \emph{\bibinfo{journal}{Phys. Rev. Mater.}} \textbf{\bibinfo{volume}{3}}, \bibinfo{pages}{064405} (\bibinfo{year}{2019}).
\newblock \urlprefix\url{https://doi.org/10.1103/physrevmaterials.3.064405}.

\bibitem{Basini2024}
\bibinfo{author}{Basini, M.} \emph{et~al.}
\newblock \bibinfo{title}{Terahertz electric-field-driven dynamical multiferroicity in srtio3}.
\newblock \emph{\bibinfo{journal}{Nature}} \textbf{\bibinfo{volume}{628}}, \bibinfo{pages}{534--539} (\bibinfo{year}{2024}).
\newblock \urlprefix\url{https://doi.org/10.1038/s41586-024-07175-9}.

\bibitem{Saparov2022}
\bibinfo{author}{Saparov, D.}, \bibinfo{author}{Xiong, B.}, \bibinfo{author}{Ren, Y.} \& \bibinfo{author}{Niu, Q.}
\newblock \bibinfo{title}{Lattice dynamics with molecular berry curvature: Chiral optical phonons}.
\newblock \emph{\bibinfo{journal}{Phys. Rev. B}} \textbf{\bibinfo{volume}{105}}, \bibinfo{pages}{064303} (\bibinfo{year}{2022}).
\newblock \urlprefix\url{https://doi.org/10.1103/PhysRevB.105.064303}.

\bibitem{Ueda2023}
\bibinfo{author}{Ueda, H.} \emph{et~al.}
\newblock \bibinfo{title}{Chiral phonons in quartz probed by x-rays}.
\newblock \emph{\bibinfo{journal}{Nature}} \textbf{\bibinfo{volume}{618}}, \bibinfo{pages}{946--950} (\bibinfo{year}{2023}).
\newblock \urlprefix\url{https://www.nature.com/articles/s41586-023-06016-5}.

\bibitem{Vonsovskii1962}
\bibinfo{author}{Vonsovskii, V.} \& \bibinfo{author}{Svirki, M.}
\newblock \bibinfo{title}{Phonon spin}.
\newblock \emph{\bibinfo{journal}{Sov. Phys. Solid State}} \textbf{\bibinfo{volume}{3}}, \bibinfo{pages}{1568--1570} (\bibinfo{year}{1962}).
\newblock \urlprefix\url{https://jglobal.jst.go.jp/en/detail?JGLOBAL_ID=201602006145839152}.

\bibitem{Levine1962}
\bibinfo{author}{Levine, A.~T.}
\newblock \bibinfo{title}{A note concerning the spin of the phonon}.
\newblock \emph{\bibinfo{journal}{Il Nuovo Cimento}} \textbf{\bibinfo{volume}{26}}, \bibinfo{pages}{190--193} (\bibinfo{year}{1962}).
\newblock \urlprefix\url{https://doi.org/10.1007/BF02754355}.

\bibitem{ueda2025_nc}
\bibinfo{author}{Ueda, H.} \emph{et~al.}
\newblock \bibinfo{title}{Chiral phonons in polar linbo$_{3}$}.
\newblock \emph{\bibinfo{journal}{Nature Communications}} \textbf{\bibinfo{volume}{17}}, \bibinfo{pages}{212} (\bibinfo{year}{2025}).
\newblock \urlprefix\url{https://doi.org/10.1038/s41467-025-66911-5}.

\bibitem{Geilhufe2023}
\bibinfo{author}{Geilhufe, R.~M.} \& \bibinfo{author}{Hergert, W.}
\newblock \bibinfo{title}{Electron magnetic moment of transient chiral phonons in ${\mathrm{ktao}}_{3}$}.
\newblock \emph{\bibinfo{journal}{Phys. Rev. B}} \textbf{\bibinfo{volume}{107}}, \bibinfo{pages}{L020406} (\bibinfo{year}{2023}).
\newblock \urlprefix\url{https://doi.org/10.1103/PhysRevB.107.L020406}.

\bibitem{Chen2025a}
\bibinfo{author}{Chen, H.} \emph{et~al.}
\newblock \bibinfo{title}{Ultrafast switching of photoinduced phonon chirality in the antiferrochiral bpo crystal} (\bibinfo{year}{2025}).
\newblock \urlprefix\url{https://doi.org/10.48550/arXiv.2506.03742}.

\bibitem{Hernandez2023}
\bibinfo{author}{Hernandez, F. G.~G.} \emph{et~al.}
\newblock \bibinfo{title}{Observation of interplay between phonon chirality and electronic band topology}.
\newblock \emph{\bibinfo{journal}{Science Advances}} \textbf{\bibinfo{volume}{9}}, \bibinfo{pages}{eadj4074} (\bibinfo{year}{2023}).
\newblock \urlprefix\url{https://www.science.org/doi/abs/10.1126/sciadv.adj4074}.

\bibitem{Dornes2019}
\bibinfo{author}{Dornes, C.} \emph{et~al.}
\newblock \bibinfo{title}{The ultrafast einstein--de haas effect}.
\newblock \emph{\bibinfo{journal}{Nature}} \textbf{\bibinfo{volume}{565}}, \bibinfo{pages}{209--212} (\bibinfo{year}{2019}).
\newblock \urlprefix\url{https://doi.org/10.1038/s41586-018-0822-7}.

\bibitem{Tauchert2022}
\bibinfo{author}{Tauchert, S.~R.} \emph{et~al.}
\newblock \bibinfo{title}{Polarized phonons carry angular momentum in ultrafast demagnetization}.
\newblock \emph{\bibinfo{journal}{Nature}} \textbf{\bibinfo{volume}{602}}, \bibinfo{pages}{73--77} (\bibinfo{year}{2022}).
\newblock \urlprefix\url{https://doi.org/10.1038/s41586-021-04306-4}.

\bibitem{Olaniyan2024}
\bibinfo{author}{Olaniyan, I.} \emph{et~al.}
\newblock \bibinfo{title}{Switchable topological polar states in epitaxial batio3 nanoislands on silicon}.
\newblock \emph{\bibinfo{journal}{Nature Communications}} \textbf{\bibinfo{volume}{15}}, \bibinfo{pages}{10047} (\bibinfo{year}{2024}).
\newblock \urlprefix\url{https://doi.org/10.1038/s41467-024-54285-z}.

\bibitem{Cohen1992}
\bibinfo{author}{Cohen, R.~E.}
\newblock \bibinfo{title}{Origin of ferroelectricity in perovskite oxides}.
\newblock \emph{\bibinfo{journal}{Nature}} \textbf{\bibinfo{volume}{358}}, \bibinfo{pages}{136--138} (\bibinfo{year}{1992}).
\newblock \urlprefix\url{https://doi.org/10.1038/358136a0}.

\bibitem{Moseni2022}
\bibinfo{author}{Moseni, K.}, \bibinfo{author}{Wilson, R.~B.} \& \bibinfo{author}{Coh, S.}
\newblock \bibinfo{title}{Electric field control of phonon angular momentum in perovskite ${\mathrm{batio}}_{3}$}.
\newblock \emph{\bibinfo{journal}{Phys. Rev. Mater.}} \textbf{\bibinfo{volume}{6}}, \bibinfo{pages}{104410} (\bibinfo{year}{2022}).
\newblock \urlprefix\url{https://link.aps.org/doi/10.1103/PhysRevMaterials.6.104410}.

\bibitem{Yoo2025}
\bibinfo{author}{Yoo, S.} \emph{et~al.}
\newblock \bibinfo{title}{Ferroelectric transistors for low-power nand flash memory}.
\newblock \emph{\bibinfo{journal}{Nature}} \textbf{\bibinfo{volume}{648}}, \bibinfo{pages}{320--326} (\bibinfo{year}{2025}).
\newblock \urlprefix\url{https://doi.org/10.1038/s41586-025-09793-3}.

\bibitem{Zhang2025_sci}
\bibinfo{author}{Zhang, X.} \emph{et~al.}
\newblock \bibinfo{title}{Single-crystalline batio3-based ferroelectric capacitive memory via membrane transfer}.
\newblock \emph{\bibinfo{journal}{Science Advances}} \textbf{\bibinfo{volume}{11}}, \bibinfo{pages}{eadz2553} (\bibinfo{year}{2025}).
\newblock \urlprefix\url{https://www.science.org/doi/abs/10.1126/sciadv.adz2553}.

\bibitem{Nordlander2020}
\bibinfo{author}{Nordlander, J.} \emph{et~al.}
\newblock \bibinfo{title}{Ferroelectric domain architecture and poling of ${\mathrm{batio}}_{3}$ on si}.
\newblock \emph{\bibinfo{journal}{Phys. Rev. Mater.}} \textbf{\bibinfo{volume}{4}}, \bibinfo{pages}{034406} (\bibinfo{year}{2020}).
\newblock \urlprefix\url{https://doi.org/10.1103/PhysRevMaterials.4.034406}.

\bibitem{yang2025}
\bibinfo{author}{Zhang, X.} \emph{et~al.}
\newblock \bibinfo{title}{Catalogue of chiral phonon materials}.
\newblock \emph{\bibinfo{journal}{arXiv}}  (\bibinfo{year}{2025}).
\newblock \urlprefix\url{https://doi.org/10.48550/arXiv.2506.13721}.

\bibitem{Chiu2022}
\bibinfo{author}{Chiu, C.-C.} \emph{et~al.}
\newblock \bibinfo{title}{Presence of delocalized ti 3d electrons in ultrathin single-crystal srtio3}.
\newblock \emph{\bibinfo{journal}{Nano Letters}} \textbf{\bibinfo{volume}{22}}, \bibinfo{pages}{1580--1586} (\bibinfo{year}{2022}).
\newblock \urlprefix\url{https://doi.org/10.1021/acs.nanolett.1c04434}.

\bibitem{Leroy2025}
\bibinfo{author}{Leroy, L.} \emph{et~al.}
\newblock \bibinfo{title}{Antiferrodistortive and ferroeletric phase transitions in freestanding films of srtio3}.
\newblock \emph{\bibinfo{journal}{Nano Letters}} \textbf{\bibinfo{volume}{25}}, \bibinfo{pages}{7651--7657} (\bibinfo{year}{2025}).
\newblock \urlprefix\url{https://doi.org/10.1021/acs.nanolett.4c05664}.

\bibitem{Ament2011}
\bibinfo{author}{Ament, L.~J.}, \bibinfo{author}{Veenendaal, M.~V.}, \bibinfo{author}{Devereaux, T.~P.}, \bibinfo{author}{Hill, J.~P.} \& \bibinfo{author}{Brink, J. V.~D.}
\newblock \bibinfo{title}{Resonant inelastic x-ray scattering studies of elementary excitations}.
\newblock \emph{\bibinfo{journal}{Rev. Mod. Phys.}} \textbf{\bibinfo{volume}{83}}, \bibinfo{pages}{705--767} (\bibinfo{year}{2011}).
\newblock \urlprefix\url{https://doi.org/10.1103/revmodphys.83.705}.

\bibitem{Fan2019}
\bibinfo{author}{Fan, W.} \emph{et~al.}
\newblock \bibinfo{title}{Evolution of element-specific electronic structures in alkaline titanates}.
\newblock \emph{\bibinfo{journal}{AIP Advances}} \textbf{\bibinfo{volume}{9}}, \bibinfo{pages}{065213} (\bibinfo{year}{2019}).
\newblock \urlprefix\url{https://doi.org/10.1063/1.5109588}.

\bibitem{Ghosez1999}
\bibinfo{author}{Ghosez, P.}, \bibinfo{author}{Cockayne, E.}, \bibinfo{author}{Waghmare, U.~V.} \& \bibinfo{author}{Rabe, K.~M.}
\newblock \bibinfo{title}{Lattice dynamics of ${\mathrm{batio}}_{3},$ ${\mathrm{pbtio}}_{3}$, and ${\mathrm{pbzro}}_{3}$: A comparative first-principles study}.
\newblock \emph{\bibinfo{journal}{Phys. Rev. B}} \textbf{\bibinfo{volume}{60}}, \bibinfo{pages}{836--843} (\bibinfo{year}{1999}).
\newblock \urlprefix\url{https://link.aps.org/doi/10.1103/PhysRevB.60.836}.

\bibitem{Bieniasz2022}
\bibinfo{author}{Bieniasz, K.}, \bibinfo{author}{Johnston, S.} \& \bibinfo{author}{Berciu, M.}
\newblock \bibinfo{title}{Theory of dispersive optical phonons in resonant inelastic x-ray scattering experiments}.
\newblock \emph{\bibinfo{journal}{Phys. Rev. B}} \textbf{\bibinfo{volume}{105}}, \bibinfo{pages}{L180302} (\bibinfo{year}{2022}).
\newblock \urlprefix\url{https://link.aps.org/doi/10.1103/PhysRevB.105.L180302}.

\bibitem{Nag2025}
\bibinfo{author}{Nag, A.} \emph{et~al.}
\newblock \bibinfo{title}{Circular dichroism in resonant inelastic x-ray scattering from birefringence in cuo}.
\newblock \emph{\bibinfo{journal}{Phys. Rev. Res.}} \textbf{\bibinfo{volume}{7}}, \bibinfo{pages}{L022047} (\bibinfo{year}{2025}).
\newblock \urlprefix\url{https://doi.org/10.1103/PhysRevResearch.7.L022047}.

\bibitem{Johnston2010}
\bibinfo{author}{Johnston, S.} \emph{et~al.}
\newblock \bibinfo{title}{Systematic study of electron-phonon coupling to oxygen modes across the cuprates}.
\newblock \emph{\bibinfo{journal}{Phys. Rev. B}} \textbf{\bibinfo{volume}{82}}, \bibinfo{pages}{064513} (\bibinfo{year}{2010}).
\newblock \urlprefix\url{https://link.aps.org/doi/10.1103/PhysRevB.82.064513}.

\bibitem{Devereaux2016}
\bibinfo{author}{Devereaux, T.~P.} \emph{et~al.}
\newblock \bibinfo{title}{Directly characterizing the relative strength and momentum dependence of electron-phonon coupling using resonant inelastic x-ray scattering}.
\newblock \emph{\bibinfo{journal}{Phys. Rev. X}} \textbf{\bibinfo{volume}{6}}, \bibinfo{pages}{041019} (\bibinfo{year}{2016}).
\newblock \urlprefix\url{https://link.aps.org/doi/10.1103/PhysRevX.6.041019}.

\bibitem{Iwasaki1971}
\bibinfo{author}{Iwasaki, H.} \& \bibinfo{author}{Sugii, K.}
\newblock \bibinfo{title}{Optical activity of ferroelectric 5pbo·3geo2 single crystals}.
\newblock \emph{\bibinfo{journal}{Applied Physics Letters}} \textbf{\bibinfo{volume}{19}}, \bibinfo{pages}{92--93} (\bibinfo{year}{1971}).
\newblock \urlprefix\url{https://doi.org/10.1063/1.1653848}.

\bibitem{Deliolanis2004}
\bibinfo{author}{Deliolanis, N.~C.} \emph{et~al.}
\newblock \bibinfo{title}{Direct measurement of the dispersion of the electrogyration coefficient of photorefractive bi12geo20 crystals}.
\newblock \emph{\bibinfo{journal}{Journal of Applied Physics}} \textbf{\bibinfo{volume}{97}}, \bibinfo{pages}{023531} (\bibinfo{year}{2004}).
\newblock \urlprefix\url{https://doi.org/10.1063/1.1828585}.

\bibitem{Xiong2022}
\bibinfo{author}{Xiong, G.}, \bibinfo{author}{Chen, H.}, \bibinfo{author}{Ma, D.} \& \bibinfo{author}{Zhang, L.}
\newblock \bibinfo{title}{Effective magnetic fields induced by chiral phonons}.
\newblock \emph{\bibinfo{journal}{Phys. Rev. B}} \textbf{\bibinfo{volume}{106}}, \bibinfo{pages}{144302} (\bibinfo{year}{2022}).
\newblock \urlprefix\url{https://doi.org/10.1103/PhysRevB.106.144302}.

\bibitem{Fransson2023}
\bibinfo{author}{Fransson, J.}
\newblock \bibinfo{title}{Chiral phonon induced spin polarization}.
\newblock \emph{\bibinfo{journal}{Phys. Rev. Res.}} \textbf{\bibinfo{volume}{5}}, \bibinfo{pages}{L022039} (\bibinfo{year}{2023}).
\newblock \urlprefix\url{https://doi.org/10.1103/PhysRevResearch.5.L022039}.

\bibitem{Yaremkevich2023}
\bibinfo{author}{Yaremkevich, D.~D.} \emph{et~al.}
\newblock \bibinfo{title}{On-chip phonon-magnon reservoir for neuromorphic computing}.
\newblock \emph{\bibinfo{journal}{Nature Communications}} \textbf{\bibinfo{volume}{14}}, \bibinfo{pages}{8296} (\bibinfo{year}{2023}).
\newblock \urlprefix\url{https://doi.org/10.1038/s41467-023-43891-y}.

\bibitem{Fina2011}
\bibinfo{author}{Fina, I.} \emph{et~al.}
\newblock \bibinfo{title}{Nonferroelectric contributions to the hysteresis cycles in manganite thin films: A comparative study of measurement techniques}.
\newblock \emph{\bibinfo{journal}{Journal of Applied Physics}} \textbf{\bibinfo{volume}{109}}, \bibinfo{pages}{074105} (\bibinfo{year}{2011}).
\newblock \urlprefix\url{https://doi.org/10.1063/1.3555098}.

\bibitem{BROOKES2018175}
\bibinfo{author}{Brookes, N.} \emph{et~al.}
\newblock \bibinfo{title}{The beamline id32 at the esrf for soft x-ray high energy resolution resonant inelastic x-ray scattering and polarisation dependent x-ray absorption spectroscopy}.
\newblock \emph{\bibinfo{journal}{Nuclear Instruments and Methods in Physics Research Section A: Accelerators, Spectrometers, Detectors and Associated Equipment}} \textbf{\bibinfo{volume}{903}}, \bibinfo{pages}{175--192} (\bibinfo{year}{2018}).
\newblock \urlprefix\url{https://www.sciencedirect.com/science/article/pii/S0168900218308234}.

\bibitem{Jones_48}
\bibinfo{author}{Jones, R.~C.}
\newblock \bibinfo{title}{A new calculus for the treatment of optical systems. vii. properties of the n-matrices}.
\newblock \emph{\bibinfo{journal}{J. Opt. Soc. Am.}} \textbf{\bibinfo{volume}{38}}, \bibinfo{pages}{671--685} (\bibinfo{year}{1948}).
\newblock \urlprefix\url{https://opg.optica.org/abstract.cfm?URI=josa-38-8-671}.

\bibitem{Verstraete2025}
\bibinfo{author}{Verstraete, M.~J.} \emph{et~al.}
\newblock \bibinfo{title}{Abinit 2025: New capabilities for the predictive modeling of solids and nanomaterials}.
\newblock \emph{\bibinfo{journal}{The Journal of Chemical Physics}} \textbf{\bibinfo{volume}{163}}, \bibinfo{pages}{164126} (\bibinfo{year}{2025}).
\newblock \urlprefix\url{https://doi.org/10.1063/5.0288278}.

\bibitem{Culbertson2020}
\bibinfo{author}{Culbertson, C.~M.} \emph{et~al.}
\newblock \bibinfo{title}{Neutron total scattering studies of group ii titanates (atio3, a2+ = mg, ca, sr, ba)}.
\newblock \emph{\bibinfo{journal}{Scientific Reports}} \textbf{\bibinfo{volume}{10}}, \bibinfo{pages}{3729} (\bibinfo{year}{2020}).
\newblock \urlprefix\url{https://doi.org/10.1038/s41598-020-60475-8}.

\bibitem{bto_grimes}
\bibinfo{author}{Grimes, M.}, \bibinfo{author}{Ueda, H.}, \bibinfo{author}{Allington, C.}, \bibinfo{author}{Kummer, K.} \& \bibinfo{author}{Staub, U.}
\newblock \bibinfo{title}{Switching of chiral phonons in ferroelectrics [dataset]. european synchrotron radiation facility.} (\bibinfo{year}{2028}).
\newblock \urlprefix\url{https://doi.org/10.15151/ESRF-ES-2166894199}.

\end{thebibliography}
\end{document}